\documentclass[article]{aa} % for a article version
\usepackage{graphicx}

\newcommand\simlt{\lower.5ex\hbox{$\; \buildrel < \over \sim \;$}}
\newcommand\simgt{\lower.5ex\hbox{$\; \buildrel > \over \sim \;$}}
\newcommand\kms{km\,s$^{-1}$}

\newcommand\LL{{\cal L}}
\newcommand\halpha{H$\alpha$\,\,}
\newcommand\rs[1]{_\mathrm{#1}}
\newcommand\zr{\rs{o}}

\begin{document}

\title{Pulsar bow-shock nebulae}

\subtitle{II. Hydrodynamical Simulation}

\author{
N. Bucciantini
}
\offprints{N.Bucciantini\\
 e-mail: niccolo@arcetri.astro.it}

\institute{Dip. di Astronomia e Scienza dello Spazio,
  Universit\`a di Firenze, Largo E.Fermi 5, I-50125 Firenze, Italy\\
\email{niccolo@arcetri.astro.it}
}

\date{Received 14 November 2001 / Accepted 19 March 2002}
\abstract{
We present hydrodynamical simulations, using a 2-D two component model (ambient medium and pulsar wind have different specific heat ratios), of bow shocks in a representative regime for pulsar wind driven bow-shock nebulae. We also investigate the behaviour of a passive toroidal magnetic field wound around the pulsar velocity direction. Moreover we estimate the opacity of the bow-shock to penetration of ISM neutral hydrogen: this quantity affects observable properties of the nebula, like its size, shape, velocity and surface brightness distribution. Finally we compare these numerical results with those from an analytical model. The development of more realistic models is needed in order to tune the criteria for searches of new such objects, as well as to interpret data on the known objects.
}

\maketitle

%%%%%%%%%%%%%%%%%%%%%%%%%%%%%%%%%%%%%%%%%%%%%%%%%%%%%%%%%%%%%%%%%%%%%%%%%%%%%%%%
%%% 1
\section{Introduction}
Pulsars moving supersonically with respect to the ambient medium are expected to give rise to a bow shock. In the case of interaction with the interstellar medium (ISM) the \halpha  emission from the nebula may be detected. Although pulsars have high typical velocities, and there are more than one thousand known radio pulsars, only four bow shocks have been discovered so far: PSR~1957+20 (Kulkarni \& Hester \cite{kulkarni88}), PSR~2224+65 (Cordes et al.\ \cite{cordes93}), PSR~J0437+4715 (Bell \cite{bell95}), and PSR~0740--28 (Jones et al.\ \cite{jones01}); while it is not clear whether the cometary-like nebula associated with the isolated neutron star RX~J185635--3754 (van Kerkwijk \& Kulkarni \cite{vankerkwijk01}) is a bow shock or results from photo-ionization. In this kind of object the relativistic particles in the pulsar wind produce a synchrotron emission  too weak to be detectable in radio wavelengths (Gaensler et al.\ \cite{gaensler00}). These nebulae may be detected instead in optical Balmer lines (mostly \halpha) as a signature of a non-radiative shock moving through a partially neutral medium (Chevalier \& Raymond \cite{chevalier78}). The Balmer lines are due to the de-excitations of neutral H atoms, following collisional excitations or exciting charge-exchange processes. When recombination times are long  these are the dominant processes giving emission in the optical band, as it has been widely studied on various supernova remnants (Smith et al.\ \cite{smith91}).

A puzzling point is that two out of the four known nebulae have a shape closely matching that of a classical bow shock, while the others show a more peculiar shape with a conical tail: the nebula associated with PSR~2224+65 in fact shows a remarkably conical tip followed by a bubble (which justifies why it has been named the ``Guitar Nebula''); (Cordes \cite{cordes96}). Various hypothesis have been put forward to justify its shape: a peculiar ISM density distribution in the surroundings of the PSR~2224+65 (Cordes et al.\ \cite{cordes93}); effects like mass loading due to neutral atoms which might penetrate in the external layers of shocked material (Bucciantini \& Bandiera \cite{bucciantini01}, hereafter Paper I) . The fourth nebula, which has recently been discovered (Jones et al.\ \cite{jones01}) near PSR~0740--28, seems to represent an intermediate case, with a ``standard'' head and a conical tail.

A pulsar wind interacting with a homogeneous ISM is in principle a much easier problem to model than that with a surrounding supernova remnant, since in the latter case the ambient medium might be quite inhomogeneous. Thus our analysis will be restricted to the former case. The ISM is seen by the pulsar as a plane-parallel flow, likely with a constant density (at least on the typical length scale of the bow shock), and lasting long enough to produce a steady-state regime. However, modeling is made more complicated by the fact that the pulsar wind is relativistic and magnetized. Moreover the ISM is typically partly neutral and the neutral atoms may have collisional mean free paths comparable to the scale length of the system, then invalidating a purely fluid treatment. However as demonstrated in a previous paper (Paper I), if we suppose that the H atoms can be ionized only via collisions with electrons and protons of the shocked plasma, for a large number of pulsars the presence of a neutral component in the ISM can be neglected as far as the fluid dynamics is concerned, or at least it can be taken into account as a small perturbative effect. 

The interaction of the relativistic magnetized pulsar wind with the ISM ionized component is not direct, because the relativistic particles themselves have very small cross sections, but is mediated by the magnetic field advected by the pulsar wind and compressed on the head of the nebula. The effective mean free path of particles is then their gyroradius, which is typically much smaller than the typical dimension of the nebula, scaling with the distance of the bow shock stagnation point from the pulsar:  
\begin{equation}
  d\zr=\sqrt{\LL/4\pi c \rho\zr V\zr^{2}},
  \label{eq:bowshocksize}
\end{equation}
where $\LL$ and $V\zr$ are the pulsar luminosity and peculiar velocity, and $\rho\zr$ the density of the dragged component of ambient medium (Paper I). For this reason we have assumed little mixing between the two media, which allows the use of a  fluid approach to investigate the structure and dynamics of these objects.

In Section 2 we present the numerical code and the parameters of the various simulations we have performed; Sect.~3 discusses the results of these simulations and compares them to what is expected from an analytic ``two thin layer'' model (Comeron \& Kaper \cite{comeron98}); in Sect.~4 we evaluate the penetration thickness of the external layer for a neutral hydrogen atom of the ISM.

%%%%%%%%%%%%%%%%%%%%%%%%%%%%%%%%%%%%%%%%%%%%%%%%%%%%%%%%%%%%%%%%%%%%%%%%%%%%%%%%
%%% 2
\section{Simulation Setup and Initial Conditions}

To simulate a spherically symmetric wind from a moving star, interacting with the ambient medium, we have used the multi-D code CLAWPACK (LeVeque  \cite{leveque94}): a second order Godunov (Godunov \cite{godunov59}) with piecewise linear reconstruction and transverse flux corrections. The problem has an intrinsic cylindrical symmetry which allows us to use a two-dimensional grid with Strang Splitting (Strang \cite{strang68}) for the geometrical terms. The boundary conditions we have imposed are: 1. At every time step the values of all relevant quantities inside a sphere of given radius centered on the star are reset to the free flowing solution; 2. A uniform wind enters the computational box from the right hand side. We have put reflecting conditions on the axis of symmetry. On the left hand side of the box there is a free outflow boundary condition: all values of variables in the ghost zones are equal to the values in the corresponding active zones (extrapolating the flow beyond the boundary), taking care to avoid the possible inflow of material due to roll-up features (Stone \& Norman \cite{stone92}, LeVeque \& al. \cite{lev97}). This boundary condition, despite its simplicity, works quite well, and is actually used by various codes. We know from solar-wind simulations that the choice of such boundary conditions may influence the evolution in the case of subsonic outflow, but even in more sophisticated codes, the only way to solve the problem seems to take boundaries very far from the zone of interest (Pauls \& al. \cite{pauls95}). We have tested the code with our simple buondary, and we have found that there are no significant problems with reflected spurious waves, and even following the evolution of our simulations we do not find spurious wave. Both the stellar and the external winds are taken with high Mach numbers (hypersonic limit). In fact the pulsar wind has a thermal pressure much lower than its ram pressure and the same holds true for the ISM, because pulsars have typical peculiar velocities of 200 \kms  (Cordes \& Chernoff \cite{cordes98}) while the typical temperature of the ISM is of the order of $10^{4}-10^{5}$ K. 

The pulsar wind is highly relativistic, with a Lorentz factor up to $10^{6}$ (Rees \& Gunn \cite{rees73}; Kennel \& Coroniti \cite{kennel84}), thus in order to handle it properly  a relativistic code is required; however the geometry of such nebulae is essentially determined by the momentum flux of the winds (Wilkin \cite{wilkin96}), so even a classical model can give a reasonable approximation. We decided to use a different adiabatic coefficient for the pulsar wind material and the ambient medium, so we have modified the Roe Riemann Solver (Roe \cite{roe81}) used by the code, adding a tracer $\phi$ to separate the ambient medium (with adiabatic coefficient $\gamma_{1} = 5/3$) from  the relativistic (with $\gamma_{2} = 4/3$) fluid coming from the star (Shyue \cite{shyue98}, Karni \cite{karni94}). We then integrate the set of equations:

\begin{eqnarray}
\lefteqn{
\frac{\partial}{\partial t}\pmatrix{\rho \cr \rho v_{r} \cr \rho v_{z} \cr E \cr \rho\phi}+\frac{\partial}{\partial z}\pmatrix{\rho v_{z} \cr \rho v_{r}v_{z} \cr \rho v_{z}^{2} + P \cr \rho v_{z}H \cr \rho\phi v_{z}}+}\nonumber\\
& & \frac{\partial}{\partial r}\pmatrix{\rho v_{r} \cr \rho v_{r}^{2} + P \cr \rho v_{z}v_{r} \cr \rho v_{r}H \cr \rho\phi v_{r}}+\frac{1}{r}\pmatrix{\rho v_{r} \cr \rho v_{r}^{2} \cr \rho v_{z}v_{r} \cr \rho v_{r}H \cr \rho\phi v_{r}} = 0,
\end{eqnarray}
where $\rho$ is the density, $v_{z}$ and $v_{r}$ are the components of the velocity, $P$ $E$ and $H$ are the pressure, energy density and enthalpy of the fluid. $\phi$ takes the value $\phi_{1}$ for the ambient medium and $\phi_{2}$ for the relativistic material. So, the value of the adiabatic coefficient $\gamma$ in each cell is determined by the relation:

\begin{equation}
\frac{1}{\gamma-1}= \frac{1}{\gamma_{1}-1}\frac{\phi_{2}-\phi}{\phi_{2}-\phi_{1}} + \frac{1}{\gamma_{2}-1}\frac{\phi_{1}-\phi}{\phi_{1}-\phi_{2}}
\end{equation}

The physical conditions of such a bow-shock guarantee conservation of energy, and little mixing between the two media, so that we can use a fluid treatment. However we added a little diffusion to avoid the growth of numerical instabilities (``carbuncle'') due to the solver, and associated with grid aligned shocks. In fact these instabilities develop on the axis and affect both the bow-shock and the contact discontinuity and then move towards the tail of the nebula, leading to a complete disruption of its structure. The introduction of a little diffusion, which avoids the growth of numerical instabilities, but may also damp the physical ones, should not be considered only as a numerical trick; in fact the four known nebulae show a well-defined shape, so there must be some process (perhaps connected to the magnetic field) which tends to stabilize the flow.

\subsection{Choice of initial parameters}
 
We have performed two different types of simulations: the first one limited to the head of the nebula, using a finer grid, and the second one extended to the tail up to about 4 times the distance from the stagnation point to the star, with a coarser resolution. 

The unit length we have used, {\em Ul}, has the value $10^{16}$cm, chosen to reproduce typical pulsar bow-shock nebula dimensions (Eq.~\ref{eq:bowshocksize}). The simulations of the head were performed on a 0.40{\em Ul}$\times$0.58{\em Ul} box, using a grid of 480$\times$700 cells with the star located in (100,0). For the simulations of the tail, we used a 1.45{\em Ul}$\times$1.00{\em Ul} box with a grid of 580$\times$400 cells and the star in (420,0). The stellar wind has a velocity of 1000 \kms, a density of 0.04 cm$^{-3}$ and a Mach number of about 13 upstream of the termination shock along the bow shock axis. We have performed various simulations using different velocities for the external flow (400, 300, 200, 150 \kms) and densities varying accordingly, in order to keep the ram pressure constant. The following figures refer to the simulation of 400 \kms (density of 0.25 cm$^{-3}$). The time needed to reach the steady state condition is of the order of 4-5 times the external flow crossing time in the computational box.
%%%%%%%%%%%%%%%%%%%%%%%%%%%%%%%%%%%%%%%%%%%%%%%%%%%%%%%%%%%%%%%%%%%%%%%%%%%%%%%%
%%% 3
\section{Comparison with the analytic model}
%%%%%%%%
\subsection{Shape}
The geometry and internal structure of the bow shock nebula resulting from our simulations are shown in Fig.~1 and Fig.~2, while a grey-scale contour plot of the density is shown in Fig.~3. The positions of the termination shock, bow shock and contact discontinuity do not change among the mentioned above simulations, and also the position of the sonic point remains essentially the same. This is because the solution, in the hypersonic limit, depends only on the ratio between ram pressures of the stellar and ambient medium, but not separately on their densities or velocities: therefore we are confident that we have approximated well the hypersonic limit. The dashed lines represent three analytic solutions (Wilkin \cite{wilkin96}), one scaled to match the contact discontinuity near the head, one to match the bow-shock near the head, and the long-dashed one, which is evaluated for the correct wind/environment ram pressure ratio. It can be seen that the analytic model reproduces quite well the shape of the nebula, although an uncertainty remains in the size. Fig.~1 shows that the region of unperturbed stellar wind extends in the tail up to a ``spherical shock'', whose position is strictly connected to the value of the thermal pressure in the ambient medium. In fact in the tail the pressure tends to the values of the ambient medium $P\zr$ so the distance of the spherical shock can be easily derived as:
\begin{equation}
  d_{back}\simeq\sqrt{\LL/4\pi c P\zr}=d\zr M\zr\gamma_{1}^{1/2},
  \label{eq:backshocksize}
\end{equation}

where $M\zr$ is the external flow Mach number.

We remark that the outflow boundaries we use may introduce spurious waves, and this plays a role in the structure of the subsonic region (region A) especially as it may change the position of the triple point (where the spherical shock meets the termination shock). However the head of the simulated nebula, and the supersonic region of the external shocked layer are not affected, so that the following comparison with the analytic model is still valid.

Looking at the head (Fig.~2) we notice a bump in the contact discontinuity, due to an instability that has not been damped completely by diffusion.
The contact discontinuity in the tail (Fig.~1 and Fig.~3) tends to reach a limiting distance from the axis (i.e. the shocked pulsar wind is confined to a cylinder), with radius of the same order of the stagnation point distance; instead, the external bow-shock tends to reach the Mach cone.

An upper limit to the radius of the contact discontinuity can be evaluated using an asymptotic treatment. Asymptotically in the tail the bow shock degenerates in a Mach cone, so the pressure in the external layer is essentially the same as the thermal pressure of the ambient medium $P\zr$. Inside the contact discontinuity there are two different regions: the innermost (labeled as region A) corresponds to the material shocked in the spherical shock, which moves subsonically; the other region (region B), which surrounds the previous one, is filled with the material which crossed the termination shock at distances from the pulsar less than the spherical shock. As shown in Fig.~1 and Fig.~2, the material in the latter region moves supersonically. In a steady state model the pressure in these two regions must asymptotically match the conditions in the ambient medium. So the pressure inside the contact discontinuity is also $P\zr$ and this justifies Eq.~(\ref{eq:backshocksize}) for the position of the spherical shock.

If we consider a fluid particle moving in region B, from a position near the stagnation point (labeled with subscript $_{ini}$) towards the asymptotic region in the tail (labeled with subscript $_{asy}$), the entropy conservation requires:

\begin{equation}
P_{asy} /\rho_{asy}^{\gamma_{2}}=P_{ini}/\rho_{ini}^{\gamma_{2}}
  \label{eq:entrop}
\end{equation}

On the other hand we have: $P_{asy}=P\zr$, and:
\begin{equation}
 P_{ini}= A(\gamma_{1})\rho\zr V\zr^{2},\\
 \rho_{ini}=\rho_{ups}B(\gamma_{2}),
\end{equation}
where $\rho_{ups}$ is the value of the density upstream of the wind termination shock along the axis, and :

\begin{eqnarray}
 A(\gamma)=&
\left(\frac{\gamma+1}{2}\right)^{\frac{\gamma+1}{\gamma-1}}\left(\frac{1}{\gamma}\right)^{\frac{\gamma}{\gamma-1}} ,\nonumber\\
 B(\gamma)=&
\frac{\gamma+1}{\gamma-1}\left(\frac{\gamma+1}{2}\right)^{\frac{2}{\gamma-1}}\left(\frac{1}{\gamma}\right)^{\frac{1}{\gamma-1}} .
\end{eqnarray}

If we consider an external wind with high Mach number, the position of the spherical shock is far from the star so we can assume that all the material of the stellar wind is confined to region B and that the distance $Z$ of the contact discontinuity from the axis is much larger than the radius of internal region A. The matter flux conservation law allows us to write:

\begin{equation}
4\pi d\zr^{2}\rho_{ups}V_{wind} \simeq V_{asy}\rho_{asy}\pi Z^{2}
  \label{eq:flux}
\end{equation}

From a two thin layer model we found that $V_{asy}$, taken to be constant over the whole transverse section, is close to $V_{wind}$, the stellar wind velocity.  This can be easily understood  because, if we assume that  the tail extends to infinity (this constraint will be released below), the velocity scale for the stellar wind material, in the absence of mixing with the ambient medium, is $V_{wind}$ (in the hypersonic limit the dependence on the external flow Mach number may be neglected). More specifically from the Bernoulli equation:

\begin{equation}
\frac{V_{asy}^{2}}{2} + \frac{\gamma_{2}}{\gamma_{2}-1}\frac{P_{asy}}{\rho_{asy}}=\frac{\gamma_{2}}{\gamma_{2}-1}\frac{P_{ini}}{\rho_{ini}},
\end{equation}
together with Eq.~\ref{eq:entrop}, it follows:

\begin{equation}
V_{asy}= V_{wind}\sqrt{1-\left\{\gamma_{1}M\zr^{2}/A(\gamma_{1})\right\}^{\frac{1-\gamma_{2}}{\gamma_{2}}}}
\label{eq:vasi}
\end{equation}

Substituting $\rho_{asy}$ and $V_{asy}$ in Eq.~(\ref{eq:flux}), using respectively Eq.~(\ref{eq:entrop}) and Eq.~(\ref{eq:vasi}), we obtain:

\begin{eqnarray}
Z^{2} \simeq &  4d\zr^{2}\left\{\gamma_{1}M\zr^{2}A(\gamma_{1})\right\}^{\frac{1}{\gamma_{2}}}/\nonumber\\
& B(\gamma_{2})\sqrt{1-\left\{\gamma_{1}M\zr^{2}/A(\gamma_{1})\right\}^{\frac{1-\gamma_{2}}{\gamma_{2}}}}
\label{eq:discon}
\end{eqnarray}

This can be considered as an upper limit to the radius of the contact discontinuity. For the stellar wind material to expand sideways beyond $Z$, some process is required that reduces the velocity in the layer or increases its pressure. This can be achieved via processes like mass accretion, or if some further shock develops.
 
The distance of the bow-shock from the star given by the analytic model is 0.175 {\em Ul} (long-dashed line) while the two limiting analytic solutions of Fig.~1 and Fig.~2 (dashed lines) are at 0.22 and 0.29 {\em Ul}. So the analytic model for the correct wind/environment ram pressure ratio gives a value which corresponds to the termination shock, while the thickness of the external layer is 0.025 {\em Ul}, in good agreement with what was expected (Paper I). Even though the shocked layers of the nebula are not thin at all, we see that the analytic solution reproduces quite well the morphology, at least of the external layer, which corresponds to the \halpha emitting region. In the tail simulation we see that the contact discontinuity has not reached the asymptotic regime,  its distance from the axis being less than the maximum value given by Eq.~(\ref{eq:discon}). 

The behaviour of the fluid in regions A and B may affect the synchrotron emission from the relativistic electrons and positrons coming from the pulsar and shocked in the termination shock. As we can see from Fig.~1, material flowing in region B becomes supersonic soon in the head. In a relativistic case, the Lorentz factor associated with the bulk motion of the fluid in the tail becomes greater than the Lorentz factor associated with the thermal motion of relativistic particles. If particles have relativistic motions their synchrotron emission can be detected only if their pitch angle is close to the observer direction. In the tail the relativistic bulk motion produces a beaming of the emission in the backward direction, so that the synchrotron emission from relativistic particles, in the case of a direction of observation perpendicular to the pulsar velocity direction, can be detected only in the head, while it fades in the tail. This holds true for the less energetic particles  responsible for the radio emission and may even be true for those responsible for the X-ray emission. In region A the flow is subsonic, so the radio and X-ray emission can be detected. In this case we expect a cylindrical emitting region. If the spherical shock is far from the pulsar it may occur that the number of high energy particles confined in region A is so small that their emission might be under the threshold of detection, so that this region can be observed only in the radio band.
 
%%%%%%%%%%%%
\begin{figure}
\centering
    \resizebox{\hsize}{!}{\includegraphics{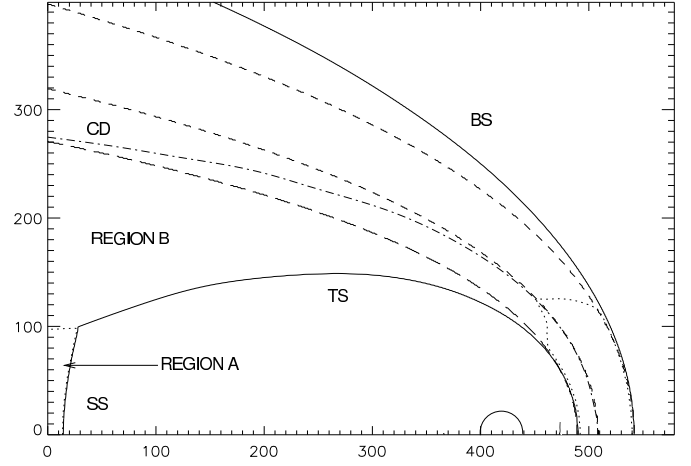}}
   \caption{Tail simulation. TS is the termination shock, SS is the spherical shock, CD is the contact discontinuity (dot-dashed line), BS is the bow shock. The short-dashed lines represent two analytic solutions, the long-dashed line is the analytic solution for the correct wind/environment ram pressure ratio, the dotted lines indicates the sonic surfaces (the subsonic region is in the head). The circle corresponds to the region in which the stellar wind values are fixed every time step.}
   \label{fig:1}
\end{figure}
%%%%%%%%%%%%
%%%%%%%%%%%%
\begin{figure}
\centering
    \resizebox{\hsize}{!}{\includegraphics{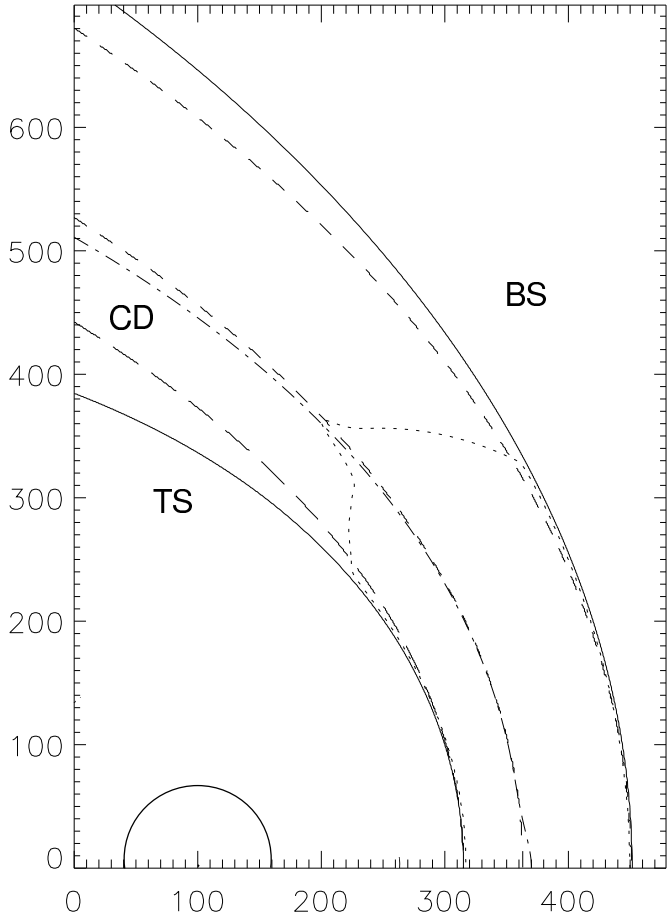}}
   \caption{Head simulation. All curves and labels are defined as in Fig.~1.}
   \label{fig:2}
\end{figure}
%%%%%%%%%%%%
%%%%%%%%%%%%
\begin{figure}
\centering
    \resizebox{\hsize}{!}{\includegraphics{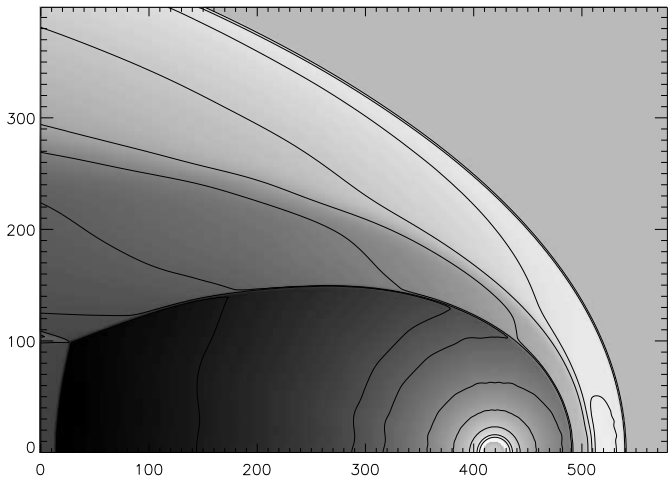}}
   \caption{Tail simulation. Density contours and shades in logarithmic scale are shown. The little spreading of the contact discontinuity is due to the diffusion we have introduced to avoid the growth of numerical instabilities.}
   \label{fig:3}
\end{figure}
%%%%%%%%%%%%
\subsection{Internal magnetic field}

Using the numerical hydrodynamic solutions described above we have also evaluated the behaviour of a toroidal magnetic field wound around the symmetry axis and passively advected by the fluid. We considered the case of a pulsar wind with the energy associated with the ordered magnetic field much smaller than the ram pressure of the relativistic particles, and with the ratio between the two being independent of the direction.  The value of such a field increases in the head, due to the compression in the termination shock, ranging from 7 times its upstream value, right after the shock, up to a factor of 30-35 with respect to its upstream value, near the contact discontinuity. In the tail it tends to decrease with respect to its value in the head due to the subsequent expansion of the fluid until it approaches a constant value:

\begin{eqnarray}
B_{asy} \simeq & \frac{\pi}{2}B_{ups}\left\{ \gamma_{1}M\zr^{2}A(\gamma_{1})\right\}^{\frac{-1}{2\gamma_{2}}}B(\gamma_{2})^{\frac{-1}{2}}\times\nonumber\\
& \left\{1-\left(\gamma_{1}M\zr^{2}/A(\gamma_{1})\right)^{\frac{1-\gamma_{2}}{\gamma_{2}}}\right\}^{-1/4}
\end{eqnarray}

In fact the contact discontinuity tends to a limit section and all the magnetic field produced by the star is confined in such region. If the original magnetic field is not too far below equipartition, it can dominate the fluid dynamics both in the tail and in the head. So an hydrodynamical approach is asymptotically valid if:

\begin{equation}
\beta_{asy}=\frac{8\pi P\zr}{B_{asy}^{2}} \gg 1
\end{equation}

Following a procedure analogous to the one used to derive the asymptotic axial distance of the contact discontinuity we are able to relate the asymptotic condition to the initial values in the head:

\begin{eqnarray}
\beta_{asy} \simeq & \left(\frac{2}{\pi}\right)^{2}\beta_{ini}\left\{\gamma_{1}M\zr^{2}A(\gamma_{1})\right\}^{\frac{1-\gamma_{2}}{\gamma_{2}}}B(\gamma_{2})\times\nonumber\\
& \left\{1-\left(\gamma_{1}M\zr^{2}/A(\gamma_{1})\right)^{\frac{1-\gamma_{2}}{\gamma_{2}}}\right\}^{1/2}
\end{eqnarray}

So, even if initially the magnetic field's energy is below equipartition, it can reach equipartition if the external wind Mach number is high enough.
Full MHD simulations are under development.

%%%%%%%%%%%%
\subsection{Surface density and tangential velocity in the external layer}

To verify that the two thin layer model could be taken as a reasonable approximation of the true structure of the nebula, at least as far as the external layer is concerned, and to evaluate how far in the tail it could be used, we have extracted from our simulations the surface density and the tangential velocity in the external layer, between the bow shock and the contact discontinuity.

Fig.~4 compares the surface density in the external layer, extracted from the tail simulation, with that evaluated by a two thin layer model. The numerical values have been obtained by integrating in the direction perpendicular to the bow shock. The difficulty in determining how good the analytic model approximation is follows from the fact that the analytic model depends on just one parameter, while the real problem has intrinsically two length scales (the stagnation point distance and the thickness in the head).  In fact we have found two limit solutions (see Fig.~1 and Fig.~2), so that we could put only an upper and a lower limit to the length scale that should be used in the analytic model. The analytic curve that best reproduces the numerical points gives an external density of 0.26 cm$^{-3}$ (instead of the true value 0.25 cm$^{-3}$)  and a stagnation point distance of 0.235 {\em Ul}, compatible with the range of the two limit solutions. The fluctuations of the numerical points are caused by round-off errors and uncertainties in the evaluation of the proper orthogonal direction, due to small oscillations in the shape of the bow shock, caused by discretisation errors in the evaluation of bow-shock position, eventually enhanced by the interpolation routine we use to find the orthogonal direction. In the simulation of the head we have noticed that, near the axis, the density values are lower than expected, and this effect corresponds to the bump of the contact discontinuity, that we have previously discussed. If corrected for the presence of the bump also in this zone we find reasonable values for the density.

%%%%%%%%%%%%
\begin{figure}
\centering
    \resizebox{\hsize}{!}{\includegraphics{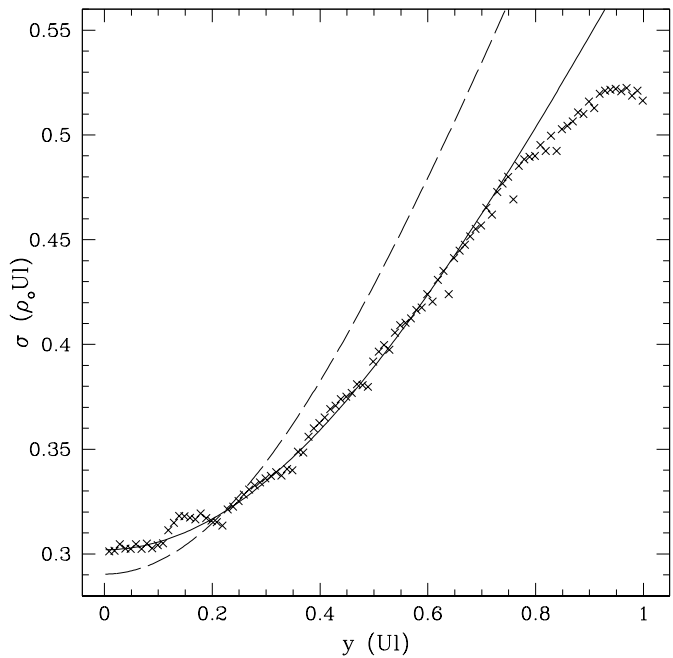}}
   \caption{Comparison of the surface density in the external layer, obtained from the tail simulation (crosses), with the values obtained from an analytic best-fit model (continuous line), and those from  the analytic solution for the correct wind/environment ram pressure ratio (dashed line).}
   \label{fig:3}
\end{figure}
%%%%%%%%%%%%

In the same way, we have evaluated the tangential velocity in the layer between the bow shock and the contact discontinuity. Fig.~5 shows the numerical values of the tangential velocity in the external layer, normalized to the external wind value, and compares it to the two thin layer model. The analytic curve that reproduces the points has been obtained using a value of 0.25 {\em Ul} for the distance of the stagnation point, a value in agreement with the two limit solutions. The fluctuations in the numerical points are much smaller than those seen in the surface density, which implies that the velocity pattern is more uniform in the layer with respect to the density. 

%%%%%%%%%%%%
\begin{figure}
\centering
    \resizebox{\hsize}{!}{\includegraphics{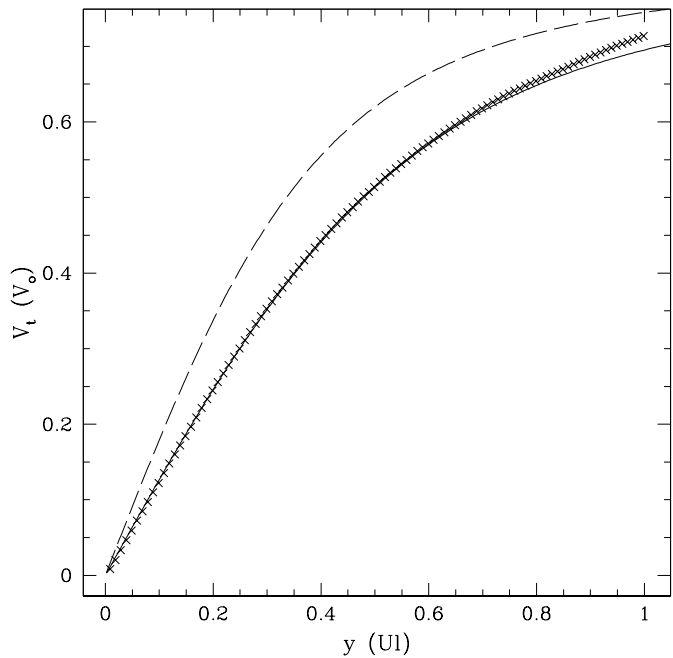}}
   \caption{Comparison of the tangential velocity in the external layer, obtained from the tail simulation (crosses), with the velocity obtained from an analytic best-fit model (continuous line), and those from  the analytic solution for the correct wind/environment ram pressure ratio (dashed line).}
   \label{fig:4}
\end{figure}
%%%%%%%%%%%%

The fact that the analytic curves that best reproduce the numerical values correspond to different values of the stagnation point distance tells us that the density and velocity cannot be reproduced exactly by the same thin layer model. It should be stressed, though, that, in spite of the non-thin structure of the external layer, the thin layer model is not a bad approximation. Deviations from the continuous curves are present only in the tail where a steady state regime may have not been achieved yet.

%%%%%%%%%%%%%%%%%%%%%%%%%%%%%%%%%%%%%%%%%%%%%%%%%%%%%%%%%%%%%%%%%%%%%%%%%%%%%%%%
%%% 4
\section{Penetration thickness for neutral hydrogen}

As we have demonstrated in a previous paper (Paper I) the penetration thickness $\tau$ for neutral hydrogen in the external layer, depends on the ambient density and the pulsar velocity.  For a pulsar moving in a partially neutral medium (the ISM ionization fraction is of order 0.1; Taylor \& Cordes \cite{taylor93}; Dickey \& Lockman \cite{dickey90}), this parameter plays an essential role in determining the true size of the nebula and the \halpha photon flux. The penetration thickness depends on the process that could drag the atoms: essentially by charge-exchange and by ionization via the interaction with the protons confined in the external layer. The cross sections of such processes are strongly dependent on velocity (McClure \cite{mcclure66}; Newman et al. \cite{newman82}; Peek \cite{peek66}; Ptak \& Stoner \cite{ptak73}), and variations of the surface density in the external layer give rise to changes of the thickness moving away from the stagnation point. In our previous paper we have evaluated the thickness using the two thin layer model limited to a region near the axis. Our aim is to verify if such an approximation is acceptable and how far in the tail can it be extended. 

The thickness of the external layer has been evaluated integrating  the ionization and charge-exchange probabilities along the path of a neutral test hydrogen atom.  So we have taken into account variations of density, relative velocity and temperature. We have verified that, for typical  pulsar velocities, collisional ionization can be neglected, even in non equilibrium cases.
 
%%%%%%%%%%%%
\begin{figure}
\centering
    \resizebox{\hsize}{!}{\includegraphics{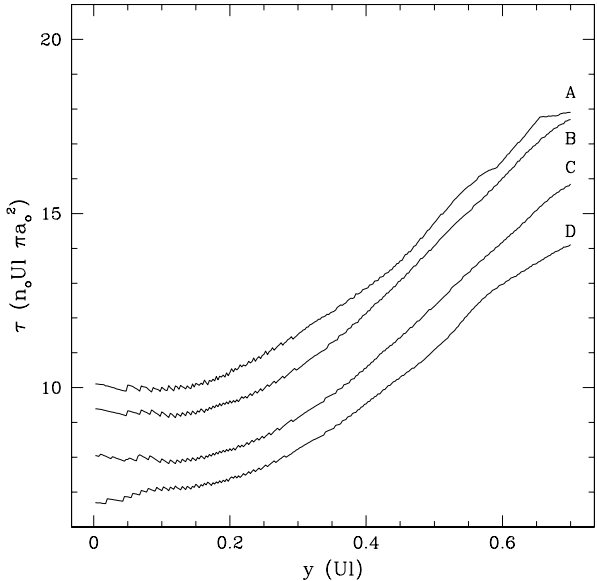}}
   \caption{Penetration thickness of the external layer for neutral hydrogen coming from the ambient medium. n$\zr$ is the ambient particle density, a$\zr$ is the Bohr Radius. The four cases represent different velocity of the pulsar: A 150 \kms; B 200 \kms; C 300 \kms; D 400 \kms.}
   \label{fig:5}
\end{figure}
%%%%%%%%%%%%

Fig.~6 shows the penetration thickness, normalized to the external density. 
The differences in the values for the various simulations are  due to variations of the cross section with velocity, while the effect of temperature is less sensitive. The thickness decreases at higher velocities as expected: in fact it follows the behaviour of the cross section for charge-exchange, which is the dominant process. Concerning the variation of $\tau$ with the distance from the stagnation point, we can see that it has a homogeneous trend for different velocity values. Moving away from the stagnation point, the surface density increases and the same holds for the cross section of charge-exchange, due to the decrease of relative velocity between the neutral hydrogen atoms coming from ISM and the protons flowing in the external layer. This effect, that tends to increase the penetration thickness, is however reduced because the interaction rate, which depends on the relative velocity, decreases even if the cross section is higher, so that the mean free path is longer. The thickness, at least for the head, ranges only over a factor 2. Thus we can say that the model restricted to the stagnation point, which we have proposed (Paper I), can reproduce within the same factor the behaviour of the whole head. This can be quite important, because, if the nebula is thin to the penetration of neutral hydrogen, it can remain thin in the tail, even far from the pulsar.
   
%%%%%%%%%%%%%%%%%%%%%%%%%%%%%%%%%%%%%%%%%%%%%%%%%%%%%%%%%%%%%%%%%%%%%%%%%%%%%%%%
%%% 5
\section{Conclusion}
We have performed hydrodynamical simulations of the interaction with an ambient homogeneous medium, of the wind of a pulsar moving supersonically. The use of a two component (the ambient medium and the pulsar wind) code allows us to verify directly if the two thin layer model gives a reasonable approximation. We have also extended our simulations to the tail. Our results show that the analytic model is a good approximation for the shape, the surface density and the tangential velocity in the external layer, and can be extended in the tail up to about 3 or 4 times the stagnation point distance from the pulsar. For the range of pulsar peculiar velocities that we have take into account, we have verified that the penetration thickness for neutral hydrogen in the head of the nebula has a rage of a factor of 2.

 This tells us that models based on the analytic two thin layer solution could be considered a reasonable starting point, over which we might introduce effects of mass loading. So it will be possible to derive a simple solution for the case in which the neutral component of the ambient medium cannot be considered as a small perturbation. The analysis we have performed gives information on the asymptotic behaviour of such nebulae. As we have pointed out in the introduction, two out of the four known objects seem not to correspond to our expectations. This is evidence that in such cases we cannot assume a purely hydrodynamical steady state condition, and we require some processes in the tail (mass loading, shocks) to model the observations. However the observational data on these objects are not good enough to allow for a clear choice of the more adequate description. Our simulations show that the penetration thickness of the nebula changes slowly moving away from the head, thus, if we assume that ionizing processes other than collisions are active, mass loading may affect the fluid motions not only in the head but also in the tail. It is known that mass loading or charge-exchange from slowly moving particles in a supersonic fluid tends to reduce the velocity and to increase the pressure. The shocked material coming from the stellar wind becomes supersonic soon in the head. In the supersonic region, mass loading due to ionization of the neutral atoms, penetrating from the ambient medium, may be due to ionizing photons coming from the hot shocked ISM, while photoionization from the pulsar may be neglected (Paper I). So mass loading may justify the presence of a conical tail, which requires an internal thermal pressure greater than what we found from purely hydrodynamical simulations, and this suggests the necessity of ionizing processes other than collisions.

We have also pointed out that the magnetic field might play an important role both in the head and in the tail, leading to a faster collimation of the flow. The presence of a component of the magnetic field wounded around the axis of the nebula may reduce the effect of mass loading preventing the flow in region B from expanding sideways and  collimating it. In fact in the case of the ``Guitar Nebula'' (the only for which we know the shape even far from the pulsar) we see that there is a conical tail near the star (which we suppose due to mass loading) which stops increasing and become a cylindrical tube. To justify this change in the shape we need some other process (that we believe could be the action of magnetic fields, which may reach equipartition) that collimates the tail. Full MHD simulations are under development, as well as relativistic HD simulations, and will be presented in a future paper.

%%%%%%%%%%%%%%%%%%%%%%%%%%%%%%%%%%%%%%%%%%%%%%%%%%%%%%%%%%%%%%%%%%%%%%%%%%%%%%%%
%%% ACKNOWLEDGEMENTS
\begin{acknowledgements}
I want to acknowledge Luca Del Zanna for the help in modifying and testing the numerical code, and Elena Amato for the fruitful discussion and the reading of this article. This work has been supported by the Italian Ministry for University and Research (MIUR) under grant Cofin2001--02--10
\end{acknowledgements}

%%%%%%%%%%%%%%%%%%%%%%%%%%%%%%%%%%%%%%%%%%%%%%%%%%%%%%%%%%%%%%%%%%%%%%%%%%%%%%%%
%%% BIB


\begin{thebibliography}{}

\bibitem[1995]{bell95}
  Bell, J. F., Bailes, M., Manchester, R. N., Weisberg, J. M., \& Lyne, A. G.
  1995, ApJ, 440, L81
\bibitem[2001]{bucciantini01} 
  Bucciantini, N., \& Bandiera, R. 2001, A\&A, 375, 1032
\bibitem[1980]{chevalier78}
  Chevalier, R. A., \& Raymond, J. C. 1978, ApJ, 225, L27
\bibitem[1980]{chevalier80}
  Chevalier, R. A., Raymond, J. C., \& Kirshner, R. P. 1980, ApJ, 235, 186
\bibitem[1998]{comeron98}
  Comeron, F., \& Kaper, L. 1998, A\&A, 338, 273
\bibitem[1996]{cordes96}
  Cordes, J. M. 1996, ``Pulsars: Problems \& Progress'', Ed. S.Johnston,
  M.A.Walker, \& M.Bailes, ASP Conf.~Series 105, 393
\bibitem[1998]{cordes98}
  Cordes, J. M., \& Chernoff, D. F. 1998, ApJ, 505, 315
\bibitem[1993]{cordes93}
  Cordes, J. M., Romani, R. W., \& Lundgren, S. C. 1993, Nature, 362, 133
\bibitem[1990]{dickey90}
  Dickey, J. M., \& Lockman, F. J. 1990, ARA\&A, 28, 215
\bibitem[2000]{gaensler00}
  Gaensler, B. M., Stappers, B. W., Frail, D. A., et al.\ 2000, MNRAS, 318, 58
\bibitem[1959]{godunov59} 
  Godunov, S. K. 1959, Mat. Sb., 47, 271
\bibitem[2001]{jones01}
  Jones, H., Stappers, B., \& Gaensler, B. 2001, The Messenger, 103, 27
\bibitem[1994]{karni94}
  Karni, S. 1994, J. Comput. Phys., 112, 31
\bibitem[1984]{kennel84}
  Kennel, C. F., \& Coroniti, F. V. 1984, ApJ, 283, 694
\bibitem[2001]{vankerkwijk01}
  van Kerkwijk, M. H., \& Kulkarni, S. R., 2001, submitted to A\&A, [astro-ph/0110065]
\bibitem[1988]{kulkarni88}
  Kulkarni, S. R., \& Hester, J. J. 1988, Nature, 335, 801
\bibitem[1994]{leveque94}
  LeVeque, R. J. 1994, CLAWPACK - a software package for solving multidimensional conservation laws. In Proceedings of the Fifth International Conference on Hyperbolic Problems: Theory, Numerics, Applications, ed. J. Glimm et al., World Scientific, June 1994. 
\bibitem[1997]{lev97}
  LeVeque, R. J., Mihalas, D., Dorfi, E. A., Muller, E., 1997, ''Computational Methods for Astrophysical Fluid Flow'', Saas-Fee Advanced Course 27, Springer-Verlag. 
\bibitem[1966]{mcclure66}
  McClure, G. W., 1966, PhRv, 148, 47
\bibitem[1982]{newman82}
  Newman, J. H., Cogan, J. D., Ziegler, D. L., et al., 1982, PhRvA, 25, 2976
\bibitem[1992]{stone92}
  Stone, J. M., Norman, M. L., 1992, ApJS, 80, 753
\bibitem[1995]{pauls95}
  Pauls, H. L., Zank, G. P., Williams, L. L., 1995, Journal of Geophysical Research, 100, 21595
\bibitem[1966]{peek66}
  Peek, J. M. 1966, PhRv, 143, 33
\bibitem[1973]{ptak73}
  Ptak, R., \& Stoner, R. E. 1973, ApJ, 185, 121
\bibitem[1973]{rees73}
  Rees, M. F., \& Gunn, F. E. 1973, MNRAS, 167, 1
\bibitem[1981]{roe81}
  Roe, P. L. 1981, J. Comput. Phys., 43, 357
\bibitem[1998]{shyue98}
  Shyue, K. M. 1998, J. Comput. Phys., 142, 208
\bibitem[1991]{smith91}
  Smith, R. C., Kirshner, R. P., Blair, W. P., \& Winkler, P. F. 1991, ApJ,
  375, 652
\bibitem[1968]{strang68}
  Strang, G. 1968, J. Num. Anal., 5, 506
\bibitem[1993]{taylor93}
  Taylor, J. H., \& Cordes, J. M. 1993, ApJ, 411, 674
\bibitem[1996]{wilkin96}
  Wilkin, F. P. 1996, ApJ, 459, L31
\end{thebibliography}
\end{document}